\documentclass[12pt,aps,roupedaddress,showkeys]{revtex4}

\usepackage{color}
\usepackage{latexsym}

\usepackage{amsmath}
\usepackage{graphicx}

\usepackage{dcolumn}

\input epsf

\def\lesssim{\mathrel{\hbox{\rlap{\hbox
{\lower4pt\hbox{$\sim$}}}\hbox{$<$}}}}
 
\def\gtrsim{\mathrel{\hbox{\rlap{\hbox
{\lower4pt\hbox{$\sim$}}}\hbox{$>$}}}}

\begin{document}

\title{Numerical Simulation of Rotating Accretion Disk Around the
Schwarzschild Black Hole Using  GRH Code }	
\author
{Orhan D\"{o}nmez\footnote{electronic address:odonmez@nigde.edu.tr}}

\address{Nigde University Faculty of Art and Science, 
Physics Department, Nigde, Turkey} 


\date{\today}

\begin{abstract}
The $2D$ time dependent solution of thin accretion disk in a close binary 
system  have been presented on the equatorial plane around the 
Schwarzschild black hole. To do that, the special part of the General
Relativistic Hydrodynamical(GRH) equations  are solved using High 
Resolution Shock Capturing (HRSC) schemes. The spiral shock waves on the 
accretion disk are modeled using perfect fluid equation of state with 
adiabatic indices $\gamma = 1.05, 1.2$ and $5/3$. The results show that 
the spiral shock waves are created for gammas except the case $\gamma=5/3$.
These results consistent with results from Newtonian hydrodynamic code 
except close to black hole. Newtonian approximation does not  give good 
solution while matter closes to black hole. Our simulations illustrate that
the spiral shock waves are created close to black hole and  the location 
of inner radius of spiral shock wave is around $10M$ and it depends on 
the specific heat rates. We also find that the smaller $\gamma$ is 
the more tightly the spiral winds.
\end{abstract}


\maketitle

\section{INTRODUCTION}
\label{introduction}
Rotating accretion disk around compact objects is an important
problem in astrophysics, such as neutron stars and black holes which
involve mass transfer from one object to another. Shock waves in a
rotating accretion disk onto compact objects transfer the gravitation
energy to the radiation energy which is observer by different X-ray
observatory satellite, such as Chandra. Fluctuations and oscillating of
radiation emitted from these systems constantly remind us of time
variations of the dynamical quantities. These variation can happen on
times scale of a few microseconds to few years. In order to understand
these kinds of events we have started doing some numerical simulation
and looking for shock waves on an accretion disk.

In binary systems in which a compact primary star accretes, through an 
accretion disk from a lobe-filling secondary star, the tidal interaction 
with the companion can results in the formation of a two-armed spiral 
structure(Dgani \emph{et al.} 1994, Godon \emph{et al.} 1997). 
These spiral waves can transport 
angular momentum in the accretion disk which has been  suggested by 
analogy to the galactic dynamics context (Rozyczka \emph{et al.} 1993).

The problem of accretion disk on to black hole has been previously
analyzed numerically by Hawley \emph{et al.} (1985) in general relativity. 
They have
also suggested shock waves on the rotating accretion disk. After that
so many astrophysicist worked on this kinds of problem using with
Newtonian(pseudo-Newtonian gravitational potential approximation is
used) and general relativistic hydrodynamics. In Chakabarti \emph{et al.} 
(1993), They
have compared the analytic  and numerical studies  of shock wave on
inviscid accretion disk flows using with the smoothed particle
hydrodynamics (SPH) code. They have believed that shocks waves could be
common  in accretion disk and standing  shock waves can only be
produced in accretion disk around the black hole. Near to black hole
density of accretion disk in the sub-Keplerian flows is higher than
density of accretion disk in Keplerian flows the consequence of the
presence of centrifugal barrier, which is smaller at sub-Keplerian
flow. Because of this high density  standing shock can form in
accretion disk. This high density flow intercepts soft photons from a
cold Keplerian disk and reprocesses them to form high energy X-rays
(Chakabarti \emph{et al.} 1997). In  Lanzafame \emph{et al.} (1998), 
they have used SPH code with viscosity to
look at the shock waves on an accretion disk at parameter range. They
have found that if viscosity parameter is less than critical values,
shock can form. If it is bigger than critical values, shock wave
disappears. In the intermediate viscosity, the disk oscillates in the
viscous time scales. Nonaxisymmetric shock waves are found on the
equatorial plane by Molteni \emph{et al.} (1999). They have used SPH 
and Eulerian type
with TVD 
codes  to look for nonaxisymmetric shock wave applying a perturbation
on an accretion disk at different parameter range, which are specific
angular  momentum and internal energy. They have concluded that shock
waves are found by Chakabarti with perturbation in a rotating inviscid
accretion flow are generally unstable to azimuthal perturbation, but
the instability are taken care of at low level  and new stable
asymmetric accretion disk is developed with a strong shock rotating
steadily.

In Makita \emph{et al.} (2000), first, they have reviewed the spiral 
shock wave in
$2D$ and $3D$ and showed their results to look for consistency with
literature. One of the main problem in accretion disk is the mechanism
of angular momentum transport. One of the reason for angular momentum
transport  is  $\alpha$-disk
model. The viscosity is supposed to transform the angular momentum and
it can be produce shock waves depends on the that parameter. The
another way of the transporting angular momentum is the spiral shock
waves. The first convincing evidence of spiral shock wave in a
accretion disk is observed by Steeghs \emph{et al.} (1997).  They used 
the technique
which is known as Doppler temography to observe spiral structure in
the accretion disk of the eclipsing dwarf nova binary IP Peg at the
outburst phase.

Here we do perturbation onto satirically symmetric steady state
accretion disk around the black
hole. Most of the numerical calculation for accretion disk are done by
Newtonian 
hydrodynamical code using relativistic approximation on it. But our
code fully inviscid general relativistic hydrodynamical   which is used
High Resolution Shock Capturing Scheme (HRSC). General relativistic
code gives us more  detail explanations when the fluid flow closes to
black hole.


\section{FORMULATION}
\label{formulation}

The GRH equations in  references
Font \emph{et al.} (2000) and Donat \emph{et al.} (1998), written 
in the standard
covariant form, consist  of the local conservation laws of the
stress-energy tensor $T^{\mu \nu }$  and the matter current density $
J^\mu$:

\begin{eqnarray}
\bigtriangledown_\mu T^{\mu \nu} = 0 ,\;\;\;\;\; 
\bigtriangledown_\mu J^\mu = 0.
\label{covariant derivative}
\end{eqnarray}

\noindent
Greek indices run from $0$ to $3$, Latin indices from $1$ to $3$, and units 
in which the speed of light $c = 1$ are used.

Defining the characteristic waves of the general
relativistic hydrodynamical equations is not trivial with imperfect
fluid stress-energy tensor. We neglect the viscosity and heat
conduction effects.  This defines the  perfect fluid
stress-energy tensor. We use this stress-energy tensor to derive the
hydrodynamical equations. With this 
perfect fluid stress-energy tensor, we can solve some problems which
are  solved by the Newtonian hydrodynamics with viscosity, such as
those involving angular momentum transport and shock waves on an
accretion disk, etc. Entropy for 
perfect fluid is conserved along the fluid lines. The stress energy tensor
for a perfect fluid is given as

\begin{equation}
T^{\mu \nu} = \rho h u^\mu u^\nu + P g^{\mu \nu}.
\label{des 7}
\end{equation}

\noindent
A perfect fluid  is a fluid that moves through spacetime with a
4-velocity $u^{\mu}$ which may vary from event to event. It exhibits a
density of mass $\rho$ and isotropic pressure $P$ in the rest frame of
each fluid element. $h$ is the specific
enthalpy, defined as

\begin{equation} 
h = 1 + \epsilon +\frac{P}{\rho}.
\label{hdot}
\end{equation}

\noindent
Here  $\epsilon$ is the specific internal energy. The equation of
state might have  the 
functional form $P = P(\rho, \epsilon)$. The perfect
gas equation of state, 

\begin{equation}
P = (\Gamma -1 ) \rho \epsilon,
\label{flux split21}
\end{equation}

\noindent
is such a functional form.

The conservation laws in the form given in Eq.(\ref{covariant
derivative}) are not suitable for 
the use in advanced numerical schemes. In order to carry out numerical
hydrodynamic evolutions such as those reported in 
Font \emph{et al.} (2000), and to
use  HRSC methods, the hydrodynamic equations after the 3+1 split
must be written as a hyperbolic system of first order flux
conservative equations. We write Eq.(\ref{covariant derivative}) in
terms of coordinate derivatives, using the coordinates ($x^0 = t, x^1,
x^2, x^3$). Eq.(\ref{covariant derivative}) is projected onto the
basis $\lbrace n^\mu, (\frac{\partial}{\partial x^i})^\mu \rbrace$,
where $n^\mu$ is a unit timelike vector normal to a given
hypersurface. After a straightforward calculation,
we get (see Font \emph{et al.} 2000),

\begin{equation}
\partial_t \vec{U} + \partial_i \vec{F}^i = \vec{S}, 
\label{desired equation}
\end{equation} 

\noindent
where $\partial_t = \partial / \partial t$ and $\partial_i = \partial
/ \partial x^i$. This basic step serves to identify the
set of unknowns, the vector of conserved quantities $\vec{U}$, and
their corresponding fluxes $\vec{F}(\vec{U})$. With the
equations in conservation form, almost every high
resolution method devised to solve hyperbolic systems of conservation
laws can be extended to GRH.

The evolved state vector $\vec{U}$ consists of  the conservative
variables $(D, S_j, \tau)$ which are conserved variables for density,
momentum and energy respectively; in terms of the
primitive variables $(\rho, v^i, \epsilon)$, this becomes 
(Font \emph{et al.} 2000)

\begin{equation}
\vec{U} = \left( 
\begin{array}{c} 
D \\
S_j \\
\tau 
\end{array} 
\right) = \left( 
\begin{array}{c}
\sqrt{\gamma} W \rho \\
\sqrt{\gamma}\rho h W^2 v_j \\
\sqrt{\gamma} (\rho h W^2 - P - W \rho) 
\end{array} 
\right). 
\label{matrix form of conserved quantities}
\end{equation}

\noindent
Here $\gamma$ is the determinant of the 3-metric $\gamma_{ij}$, $v_j$
is the fluid 3-velocity, and W is the Lorentz factor, 

\begin{equation}
W = \alpha u^0 = (1 - \gamma_{ij} v^i v^j)^{-1/2}. 
\label{Wdot1}
\end{equation}

\noindent 
The flux vectors $\vec{F^i}$ are given by \cite{FMST}

\begin{equation}
\vec{F}^i =  \left( \begin{array}{c}
\alpha (v^i - \frac{1}{\alpha} \beta^i) D \\
\alpha \lbrace (v^i - \frac{1}{\alpha} \beta^i) S_j + \sqrt{\gamma} P
\delta ^{i}_{j} \rbrace    \\
\alpha \lbrace (v^i - \frac{1}{\alpha} \beta^i) \tau + \sqrt{\gamma}
v^i P \rbrace \end{array} \right). 
\label{matrix form of Flux vector}
\end{equation}

\noindent
The spatial components of the 4-velocity $u^i$ are related to the
3-velocity by the following formula: $u^i = W (v^i - \beta^i /
\alpha)$. $\alpha$ 
and $\beta^i$ are the lapse function and the shift vector of the
spacetime respectively. The source vector $\vec{S}$ is given by
Font \emph{et al.} (2000) 

\begin{equation}
\vec{S} =  \left( \begin{array}{c}
0 \\
\alpha \sqrt{\gamma} T^{\mu \nu} g_{\nu \sigma} \Gamma^{\sigma}_{\mu j}  \\
\alpha \sqrt{\gamma} (T^{\mu 0} \partial_{\mu} \alpha - \alpha T^{\mu
\nu} \Gamma^{0}_{\mu \nu}) \end{array} \right), 
\label{matrix form of source vector}
\end{equation}

\noindent
where $\Gamma^{\alpha}_{\mu \nu}$ is the 4-dimensional Christoffel symbol

\begin{equation}
\Gamma^{\alpha}_{\mu \nu} = \frac{1}{2} g^{\alpha \beta}
(\partial_\mu g_{\nu \beta} + \partial_\nu g_{\mu \beta} -
\partial_\beta g_{\mu \nu}).
\label{Christoffer symbol}
\end{equation}

The numerical solution of general relativistic hydrodynamical equations 
and technique used are explained in detail our first paper 
D\"{o}nmez (2003) which 
gives great detail about formulations, numerical scheme, technique, 
numerical solution of GRH equations, 
Adaptive-Mesh Refinement(AMR) and solution of special relativistic test
problem.


\section{General Relativistic Hydrodynamical Test Problem}
\label{General Relativistic Hydrodynamical Test Problem}

In this section, the Schwarzschild geometry is introduced in
spherical coordinates to define sources for the general relativistic
hydrodynamical equations. Then the accretion disk
problems are numerically modeled , which have been analytically analyzed, 
to test the full GRH code in $2D$ in the equatorial plane.

\subsection{Schwarzschild Black Hole}
\label{Schwarzschild Black Hole}

The Schwarzschild solution determined by the mass $M$ gives the geometry 
in outside of a spherical  star or black hole.
The Schwarzschild spacetime metric in spherical coordinates is

\begin{eqnarray}
ds^2 = -(1 - \frac{2 M}{r}) dt^2 + (1 - \frac{2 M}{r})^{-1} dr^2 + \nonumber \\
r^2 d\theta^2 + r^2 sin^2 \theta d\phi^2 .
\label{line element polor coordinate}
\end{eqnarray}

\noindent
It behaves badly near $r=2M$; there the first term becomes zero and
the second term becomes infinite in Eq.(\ref{line element polor
coordinate}). That radius $r=2M$ is called the Schwarzschild radius  or
the Schwarzschild horizon. 

\noindent
The spacetime metric for this line element is 

\begin{eqnarray}
g_{\mu \nu} = \left(
\begin{array}{cccc}
-(1 - \frac{2 M}{r}) & 0 & 0 & 0 \\
0 & (1 - \frac{2 M}{r})^{-1} & 0 & 0 \\
0 & 0 & r^2 & 0 \\
0 & 0 & 0 & r^2 sin^2 \theta
\end{array}
\right) .
\label{g metric}
\end{eqnarray}

\noindent
The lapse function and shift vector for this metric is

\begin{eqnarray}
\beta^r = 0.0, \;\;\;  \beta^{\theta} = 0.0, \;\;\;\; \beta^{\phi} =
0.0 ,
\nonumber\\
\alpha = (1 - \frac{2 M}{r})^{1/2} .
\label{lapse and shift}
\end{eqnarray}


\subsection{The Source Terms For Schwarzschild Coordinates}
\label{Source Termes For Schwarzschild Coordinates}

The gravitational sources for the GRH equations are given by
Eq.(\ref{matrix form of source 
vector}). In order to compute the sources in Schwarzschild coordinates for
different conserved variables, Eq.(\ref{matrix form of source vector}) 
can be rewritten as,

\begin{eqnarray}
\vec{S} =  \left( 
\begin{array}{c}
0 \\
\alpha \sqrt{\gamma} T^{\mu \nu} g_{\nu \sigma} \Gamma^{\sigma}_{\mu
r}  \\
\alpha \sqrt{\gamma} T^{\mu \nu} g_{\nu \sigma} \Gamma^{\sigma}_{\mu
\theta}  \\
\alpha \sqrt{\gamma} T^{\mu \nu} g_{\nu \sigma} \Gamma^{\sigma}_{\mu
\phi}  \\
\alpha \sqrt{\gamma} (T^{\mu 0} \partial_{\mu} \alpha - \alpha T^{\mu
\nu} \Gamma^{0}_{\mu \nu}) 
\end{array} 
\right). 
\label{matrix form of source vector2}
\end{eqnarray}

\noindent
It is seen in Eq.(\ref{matrix form of source vector}) that the source for
conserved density, $D$, is zero but the other sources depend on
the components of the stress energy tensor,  Christoffel symbols, and
$4$-metric.  
After doing some straightforward calculations, the sources can be
rewritten in Schwarzschild coordinates
for each conserved variable with the following form:

\noindent
The source for the momentum equation in the radial direction is

\begin{eqnarray}
\alpha \sqrt{\gamma} T^{\mu \nu} g_{\nu \sigma} \Gamma^{\sigma}_{\mu
r} = \frac{1}{2} \alpha \sqrt{\gamma} (T^{tt} \partial_{r} g_{tt} + T^{rr} \partial_{r} g_{rr} + \nonumber \\
T^{\theta \theta} \partial_{r} g_{\theta \theta}  + T^{\phi \phi}
\partial_r g_{\phi \phi}).
\label{explicit source in radial direction}
\end{eqnarray}

\noindent
The source for the momentum equation in the $\theta$ direction is

\begin{eqnarray}
\alpha \sqrt{\gamma} T^{\mu \nu} g_{\nu \sigma} \Gamma^{\sigma}_{\mu
\theta} = \frac{1}{2} \alpha \sqrt{\gamma} T^{\phi \phi}
\partial_{\theta} g_{\phi \phi}.
\label{explicit source in theta direction}
\end{eqnarray}

\noindent
The source for the momentum equation in the $\phi$ direction is

\begin{eqnarray}
\alpha \sqrt{\gamma} T^{\mu \nu} g_{\nu \sigma} \Gamma^{\sigma}_{\mu
\phi} = 0.0.
\label{explicit source in phi direction}
\end{eqnarray}

\noindent
The source for the energy equation is

\begin{eqnarray}
\alpha \sqrt{\gamma} (T^{\mu 0} \partial_{\mu} \alpha - \alpha T^{\mu
\nu} \Gamma^{0}_{\mu \nu}) = \nonumber \\ 
\alpha \sqrt{\gamma} (T^{r t}
\partial_{r} \alpha - \alpha T^{r t} g^{tt} \partial_{r} g_{tt}).
\label{explicit source for energy equation}
\end{eqnarray}

The non-zero components of the stress-energy tensor in Schwarzschild
coordinates can be computed by Eq.(\ref{des 7});
they are 

 \begin{eqnarray}
T^{tt} = \rho h \frac{W^2}{\alpha^2} + P g^{tt} \nonumber \\
T^{rr} = \rho h W^2 (v^r)^2 + P g^{rr} \nonumber \\
T^{\theta \theta} = \rho h W^2 (v^{\theta})^2 + P g^{\theta \theta} \\ 
T^{\phi \phi} = \rho h W^2 (v^{\phi})^2 + P g^{\phi \phi} \nonumber \\ 
T^{tr} = \rho h \frac{W^2}{\alpha} v^r . \nonumber
\label{component of stres energy tensor}
\end{eqnarray}


\subsection{Geodesics Flows}
\label{Geodesics Flows}

As a general relativistic test problem, the accretion of dust 
particles onto a black hole are solved. The exact solution for 
pressureless dust  is 
given in Appendix \ref{analytic solution for free fall}. 
This problem is numerically analyzed in $2D$ in spherical
coordinates at constant $\theta = \pi/2$, which is  the equatorial
plane, so the spatial numerical domain is the
$(r, \phi)$ plane. In this calculation,$2.4 \leq r \leq 20$ and  $0 \leq \phi
\leq 2\pi$ are  used for the computational domain. The initial conditions for
all variables  are chosen to have negligible values except the outer
boundary $(r = 20 M)$ where gas is continuously injected radially 
with  the  analytic density and velocity. Throughout the calculation,
whenever values at the outer boundary are needed the analytic
values are used. The code is tun until  a steady state solution is reached,  
using outflow boundary conditions at $r = 2.4$, inflow boundary
conditions at $r = 20$ and the periodic boundary in the  $\phi$
direction. It is found that the resulting numerical solution does not
develop any angular dependence during the simulation.  

\begin{center}
\begin{figure}
\centerline{\epsfxsize=14.cm \epsfysize=14.cm
\epsffile{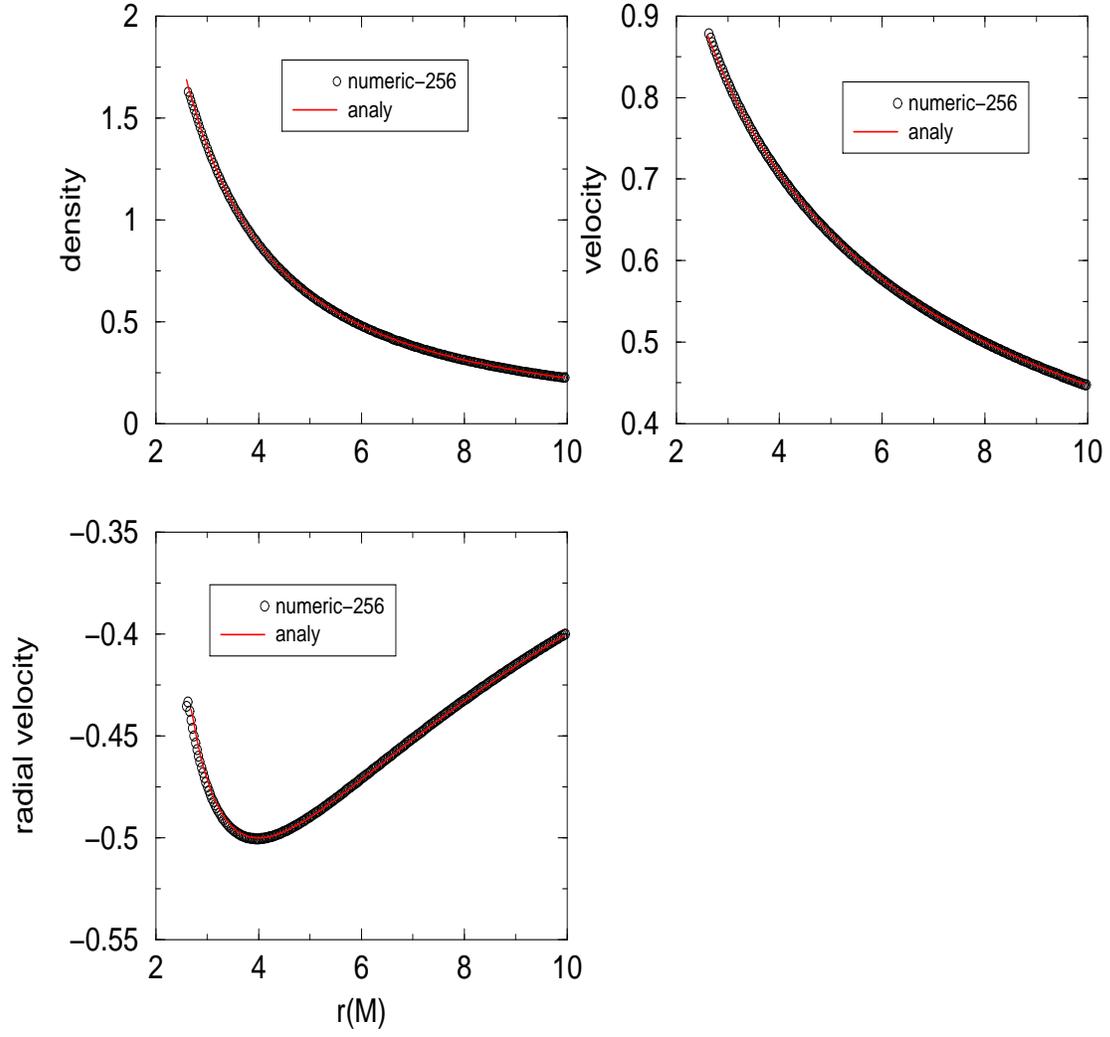}}
\caption{The analytic solutions (red solid lines)
with the numerical solutions (black circles) using $256$ zones in the
radial direction for thermodynamical variables, $\rho$, $v$, and
$v^r$.  }
\label{free falling plot for all variables}
\end{figure}
\end{center}

In Fig. \ref{free falling plot for all variables}, 
the rest-mass density $\rho$, absolute velocity $v = (v^i v_i)^{1/2}$
and radial velocity $v^r$ are plotted
as a function of radial coordinate at a fixed angular
position. The numerical solution agrees well  with the analytic
solution. The convergence test are also made on this  problem to test the
behavior of the source terms in the GR Hydro code. Some
convergence tests with the SR Hydro code are conducted to confirm
the second order convergent(D\"{o}nmez 2004). In 
this case, we are looking for convergence rate with  source terms.
The analytic values of the accretion problem are used as initial
conditions, and the computational domain is chosen from $r_{min}
= 10M$ to $r_{max}=30M$. 
The convergence results are given at 
Table\ref{convergence data for free fall}. 
It is noticed that code gives roughly second order convergence. 

The conservation form of general relativistic hydrodynamical equations 
are solved. 
It is expected that conserved variables must be conserved to machine 
accuracy, $ \sim 10^{-16}$. The checking the results of conservation variables
in the numerical test problem shows that these variables conserved to 
machine accuracy.

\begin{center}
\begin{figure}
\centerline{\epsfxsize=14.cm \epsfysize=14.cm
\epsffile{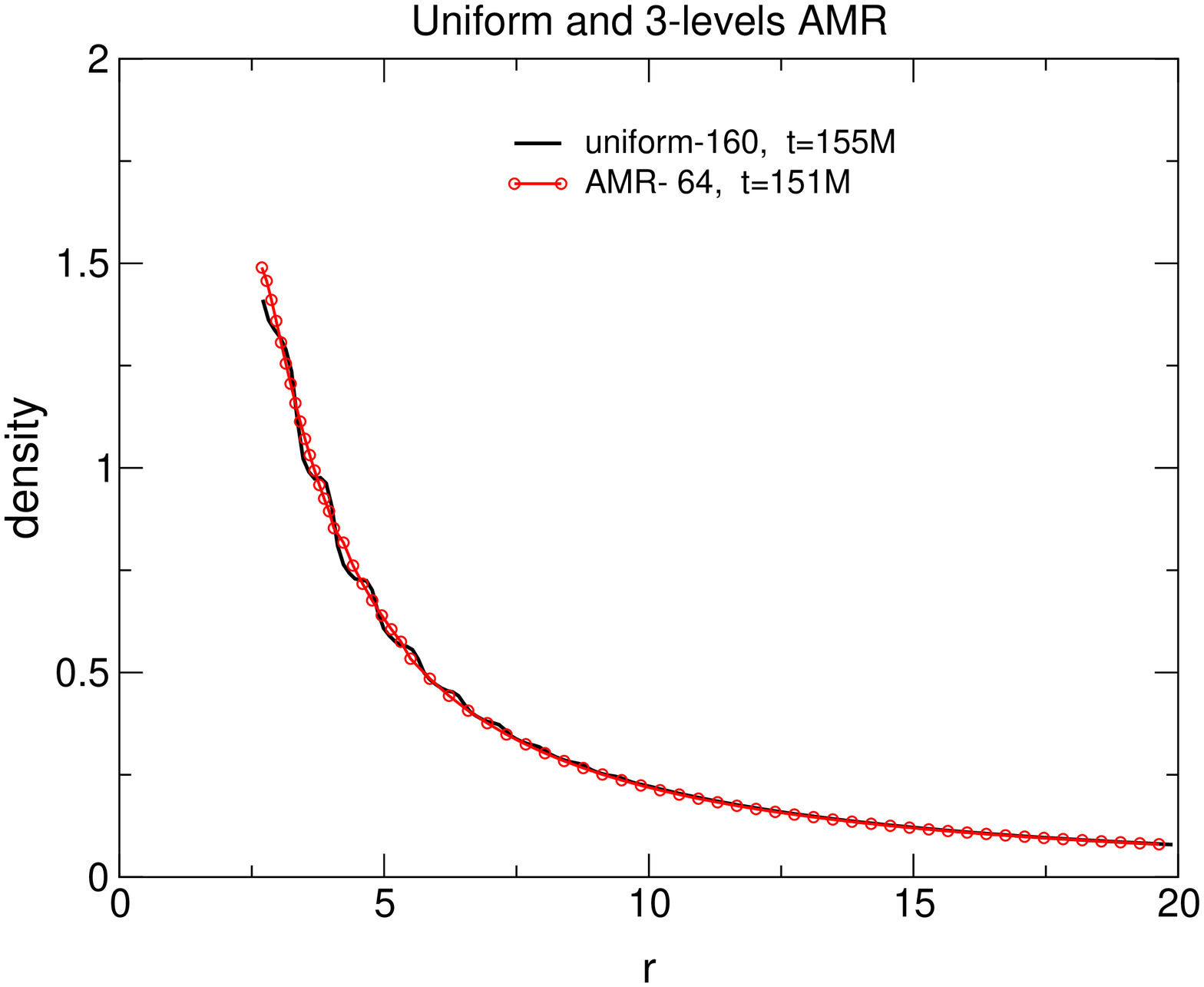}}
\caption{The numerical solutions from a $3-$level AMR run($64$
zones, red circles) and a uniform grid run($160$ zones, black straight
line) for density vs. radial coordinate.}
\label{free falling plot for AMR and Uniform}
\end{figure} 
\end{center}

In this part of same test problem we do  an AMR test.
For uniform grid runs, the amount of time it takes to reach a steady
state solution increases with resolution. We carried out a $3-$level AMR
calculation to compare with a uniform one for  
geodesic inflow  to see the behavior of AMR in this problem.
In Fig. \ref{free falling plot for AMR and Uniform} we  plot
AMR and uniform runs on top of each other for density vs. radial
Schwarzschild coordinate. We see that while the AMR solution has reached a
steady state 
at $t=151M$, the uniform solution has not reached a steady state by the
same time. 

\def\baselinestretch{1.2}

\begin{center}
\begin{table}
\caption{$L_1$ norm error and convergence factors are given  for
different resolutions.}
$$ \vbox{ \offinterlineskip \vskip8pt
  \def\qq{\hskip0.15em}
  \def\laststrut{\vrule depth7pt width0pt}
  \def\titlestrut{\vrule height12pt depth6pt width0pt}
\halign {#\vrule\strut &\quad#\hfil
       &&\qq#\vrule &\qq\hfil#\hfil \cr
\noalign{\hrule}
 & \multispan{7}\hfil Convergence Test \hfil\titlestrut 
  &\cr \noalign{\hrule}  
& \# of points &&  \# of time step && $L_1$ norm error && Convergence
  factor &\cr  
  \noalign{\hrule}
& $32$ && $1$ && $1.675E-5$ &&   \laststrut &\cr
& $64$ && $2$ && $4.0811E-6$ && 4.105 \laststrut &\cr
& $128$ && $4$ && $9.6456E-7$ && 4.23 \laststrut &\cr
& $256$ && $8$ && $2.1372E-7$ && 4.51 \laststrut &\cr
\noalign{\hrule\vskip4pt} 
  \noalign{\vskip5pt} }} $$
\label{convergence data for free fall}
\end{table} 
\end{center}

\subsection{Circular Motion of  Test Particles}
\label{Circular Motion of The Test Particle}

We will now simulate the circular motion of a fluid with the numerical code.
 To do this we set up a circular flow with negligible pressure in the 
equatorial plane, in which angular velocity at each radial direction
$r$ is the Keplerian value, Eq.(\ref{equation for circular motion9}).
This is called a Keplerian. The last stable circular orbit for a particle 
moving around a Schwarzschild black hole is at $r =6M$ ($M$ is mass of black hole) 
Therefore the gas or particles will fall into the black hole if their
radial position  is less than $6M$. When their radial position 
 is bigger than $6M$, they should rotate in a circular orbit. Here we
simulate this problem and compare the numerical solution with
the analytic expectations. This problem tests the code with sources in
the $\phi$  direction.

\begin{center}
\begin{figure}
\centerline{\epsfxsize=14.cm \epsfysize=14.cm
\epsffile{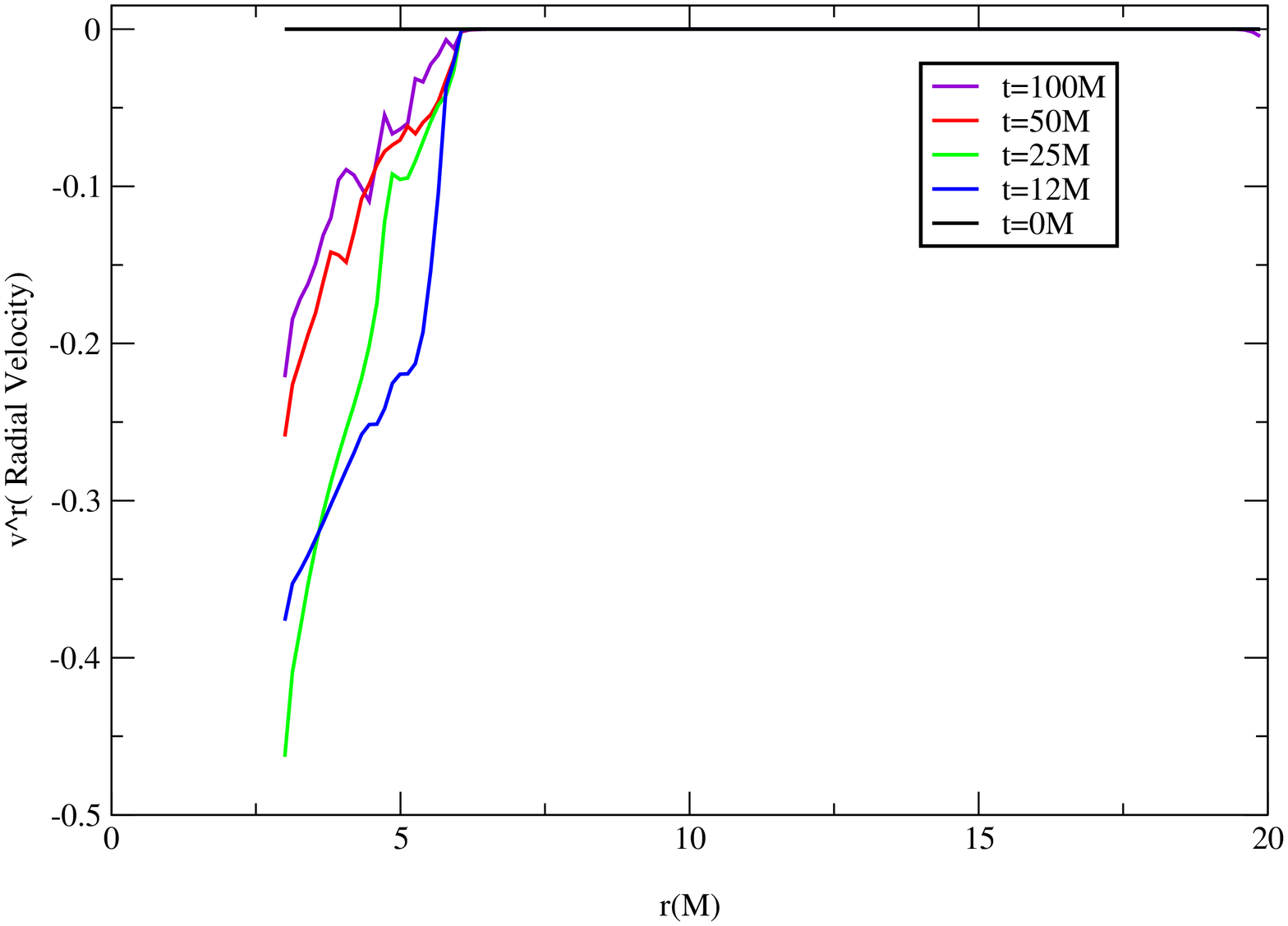}}
\caption{Radial velocity vs. $r$ is plotted. The radial velocity of fluid in the disk
stays zero for $r > 6M$, during the evolution. $r=6M$ is called the
last stable circular orbit in a Keplerian disk.}
\label{Circular motion 1}
\end{figure}
\end{center}

\begin{center}
\begin{figure}
\centerline{\epsfxsize=14.cm \epsfysize=14.cm
\epsffile{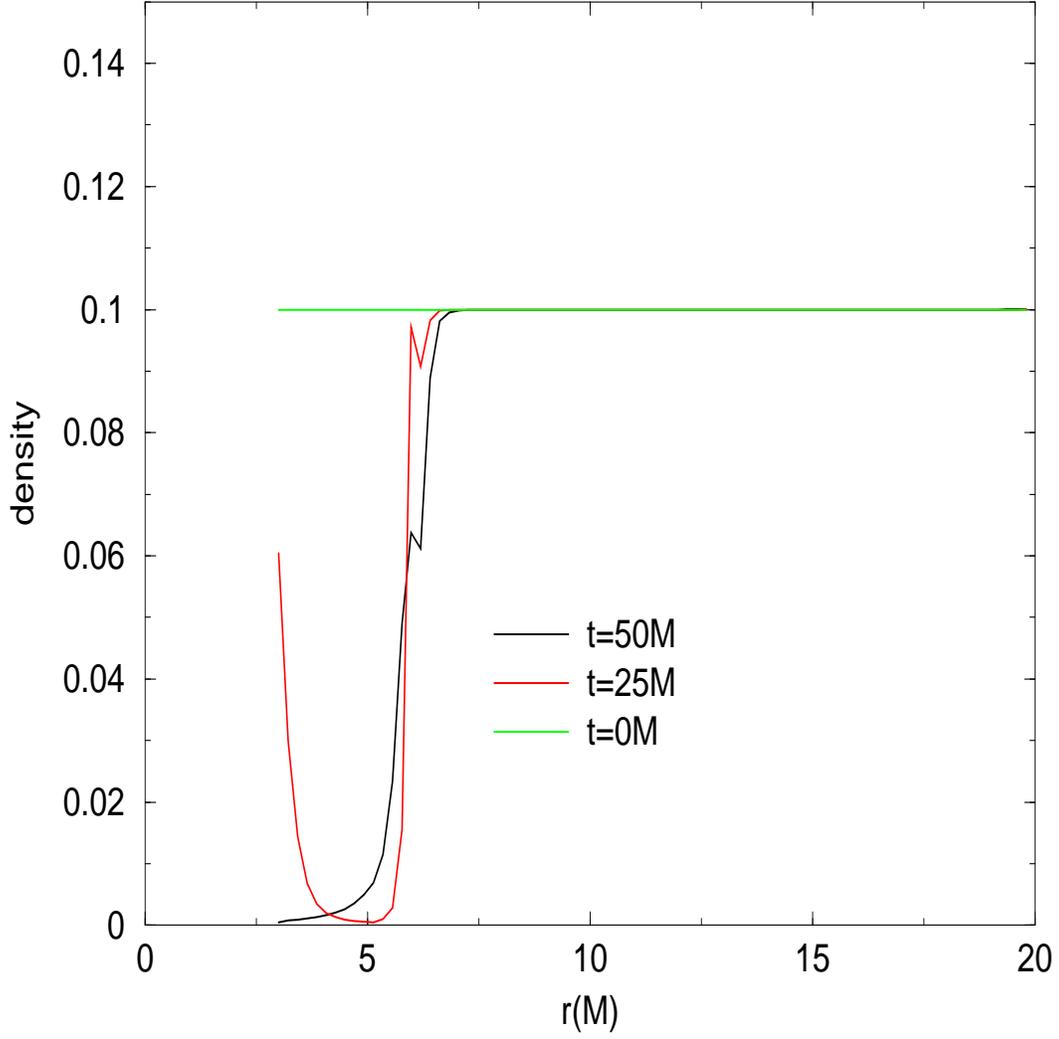}}
\caption{Density vs. $r$ is plotted. Density of the fluid is
plotted at different times using different colors. The matter falls into
the black hole while $r$ is less than $6M$.}
\label{Circular motion 2}
\end{figure}
\end{center}

In order to simulate this problem, we choose the computational domain to be 
$3M \leq r \leq 20M$ and $0 \leq \phi \leq 2 \pi$. The
computational domain is filled with constant density and pressureless gas, 
 rotating in circular orbits with the Keplerian velocity and
zero radial velocity. In Fig. \ref{Circular motion 1}, the
radial velocities of the gas vs. radial coordinate at different
times are plotted. It is numerically observed that the gas inside the 
last stable orbit, $r=6M$,
falls into the black hole while gas outside the last stable orbit follows
circular motion with the Keplerian velocity as we expect analytically. In
Fig. \ref{Circular motion 2}, we plot the density of the fluid
 vs. radial coordinate at different times to see the 
behavior of the disk.  It is clear from that gas falls into the black
hole for  $r < 6M$. So the numerical results from our code are
consistent with the analytic expectations.


\section{Numerical Modeling of the Accretion Disks in the Schwarzschild 
Coordinate}
\label{Numerical Modelling of the Accretion Disks}

In the present work, we do not intend to make a very concrete model for 
particular object, and it is  assumed a rather simple initial configuration 
of gas. At $t=0$ computational domain filled with some negligible values 
for  density and pressure  with zero radial and angular velocity. The 
injected value from companion star from outher boundary of computational 
domain are; $\rho_{in} = 1.0$, $p_{in} = 10^{-3}/\gamma$, $v^r= 0.01$ 
and $v^{\phi}$ which is the Keplerian fluid velocity. Computational 
domain is $3M \leq r \leq 100M$ and $0 \leq
\phi \leq 2\pi$. Where $M$ is the mass of black hole. Following problems 
solved are $2D$ modeling on an equatorial plane.

The choice of boundary conditions  at the inner and the  outher 
numerical boundary is rather important. The boundary conditions are treated 
as follows. The fictitious cell are placed  just outside of a boundary and we 
prescribe  physical variables in the  fictitious cells. Numerical fluxes 
on the boundary wall are computed by solving a Riemann problems between 
states. The following boundary conditions are used. The value of physical
variables in the fictitious cells are the same as those in the neighboring 
interior cells at all the times. Radial velocities  direction can be changed
depend on inner or outher boundary condition used.

\subsection{$\gamma=1.05$}
\label{gamma=1.05}
The first results of the spiral shock wave are given on the accretion 
disk around the black hole using the Schwarzschild metric. 
In all simulations, the black hole is at the center of 
computational domain and it is represented  as a white hole in the graphics.  
First, gas is injected from outher
boundary to accrete an accretion disk with spiral shock wave and then
the injection is stopped to see the behavior of accretion disk during
the evolutions. Finally the numerical simulation is stopped when solution  
reaches to steady state. 

\begin{center}
\begin{figure}
\centerline{\epsfxsize=14.cm \epsfysize=14.cm
\epsffile{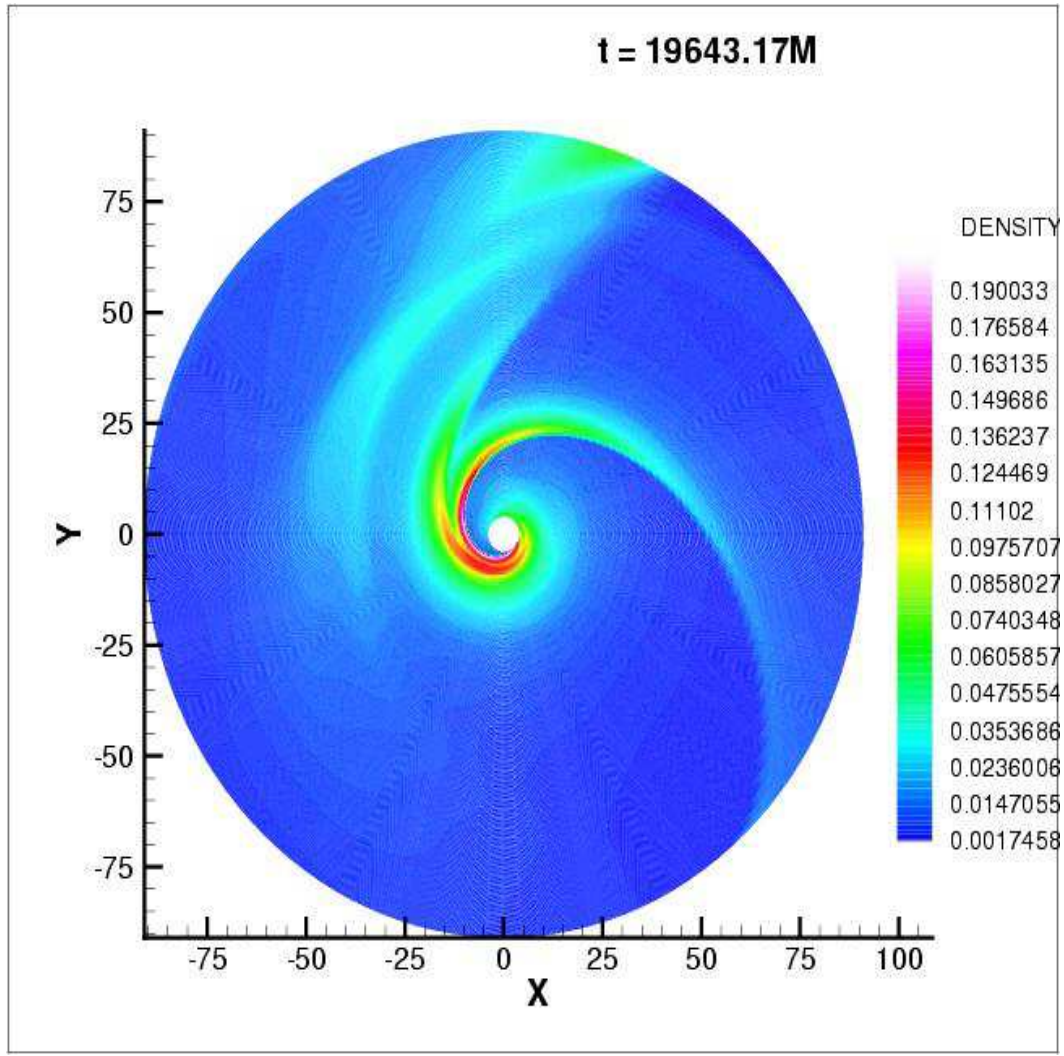}}
\caption{Plotting the density in the $r-\phi$ plane with color for
$\gamma=1.05$. It is taken at $t=19643M$ and it is in steady
state. Two-armed spiral shock wave is  created.} 
\label{gamma=1.05_2}
\end{figure}
\end{center}

\begin{center}
\begin{figure}
\centerline{\epsfxsize=14.cm \epsfysize=14.cm
\epsffile{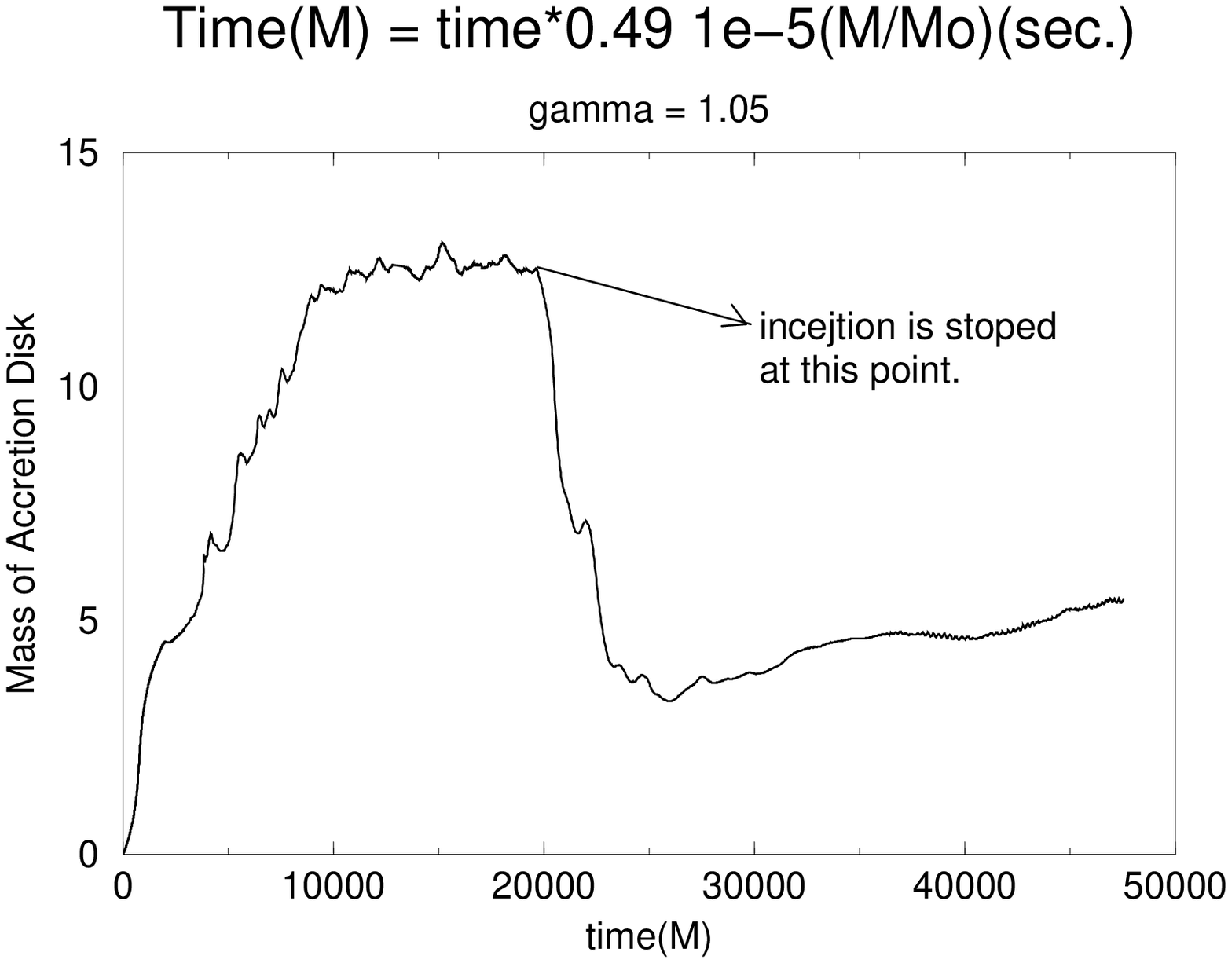}}
\caption{Mass of accretion disk vs. time is plotted during the hole
evolution. The injection is stopped at maximum mass.} 
\label{gamma=1.05_4}
\end{figure}
\end{center}

The density of the accretion disk  is illustrated in
Fig.\ref{gamma=1.05_2} at $t=19643M$. In early time of simulation 
accretion disk
is not in steady state yet but two-armed spiral shock is already created and
dynamical structure of the disk is not changed any more. The only thing
which changes during the  simulation  after the certain evolution time
is the amplitude of density. It
is also showed in Fig.\ref{gamma=1.05_4} that the mass of accretion disk during
the process of gas from outer boundary by the companion star is in
the steady states between the evolution times $10000M - 19643M$. The
two-armed spiral shock is created for the case $\gamma=1.05$ during the 
injecting gas. and also this spiral arms and accretion disk go to steady state.

\begin{center}
\begin{figure}
\centerline{\epsfxsize=14.cm \epsfysize=14.cm
\epsffile{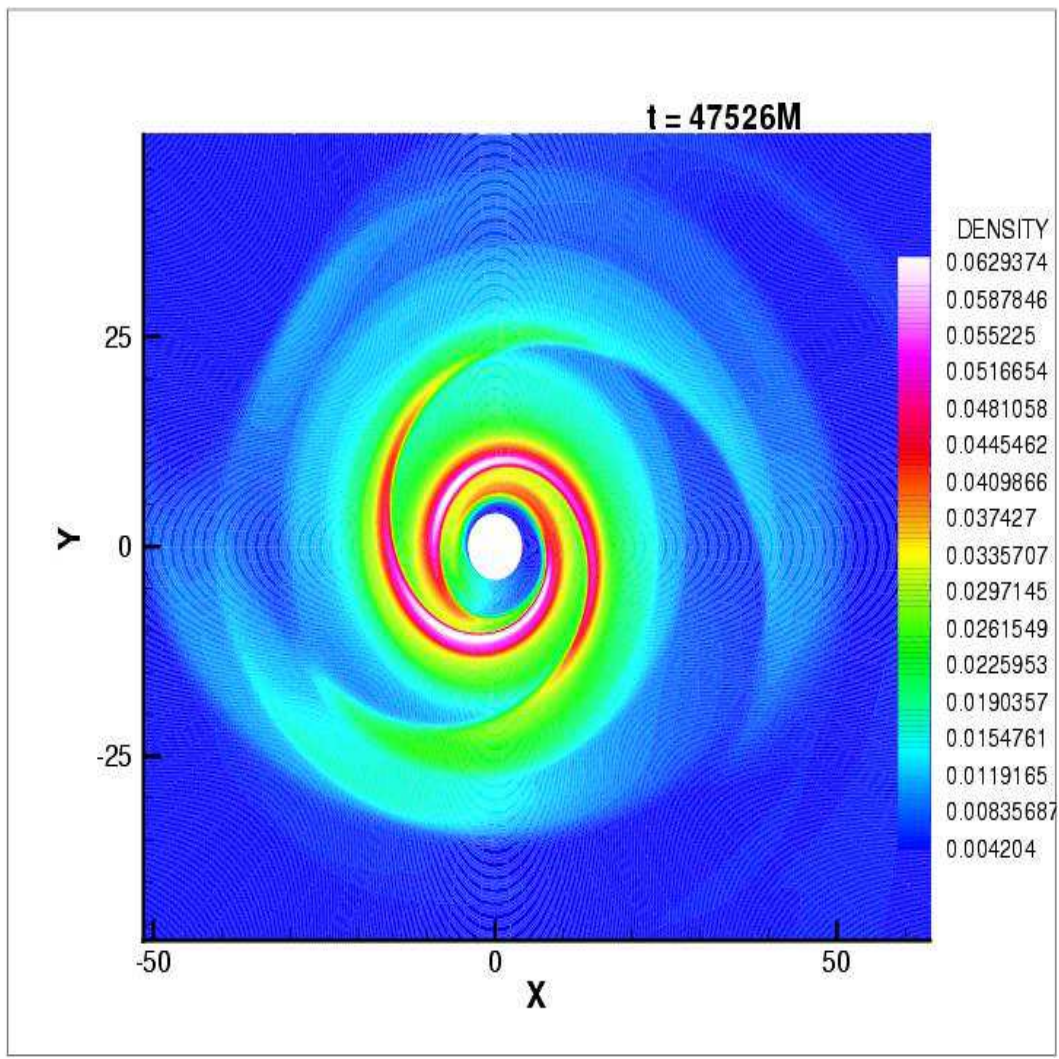}}
\caption{Zooming the interesting part of accretion disk at
$t=47526M$ to see two-armed shock wave clearly.}  
\label{gamma=1.05_5}
\end{figure}
\end{center}
\begin{center}
\begin{figure}
\centerline{\epsfxsize=14.cm \epsfysize=14.cm
\epsffile{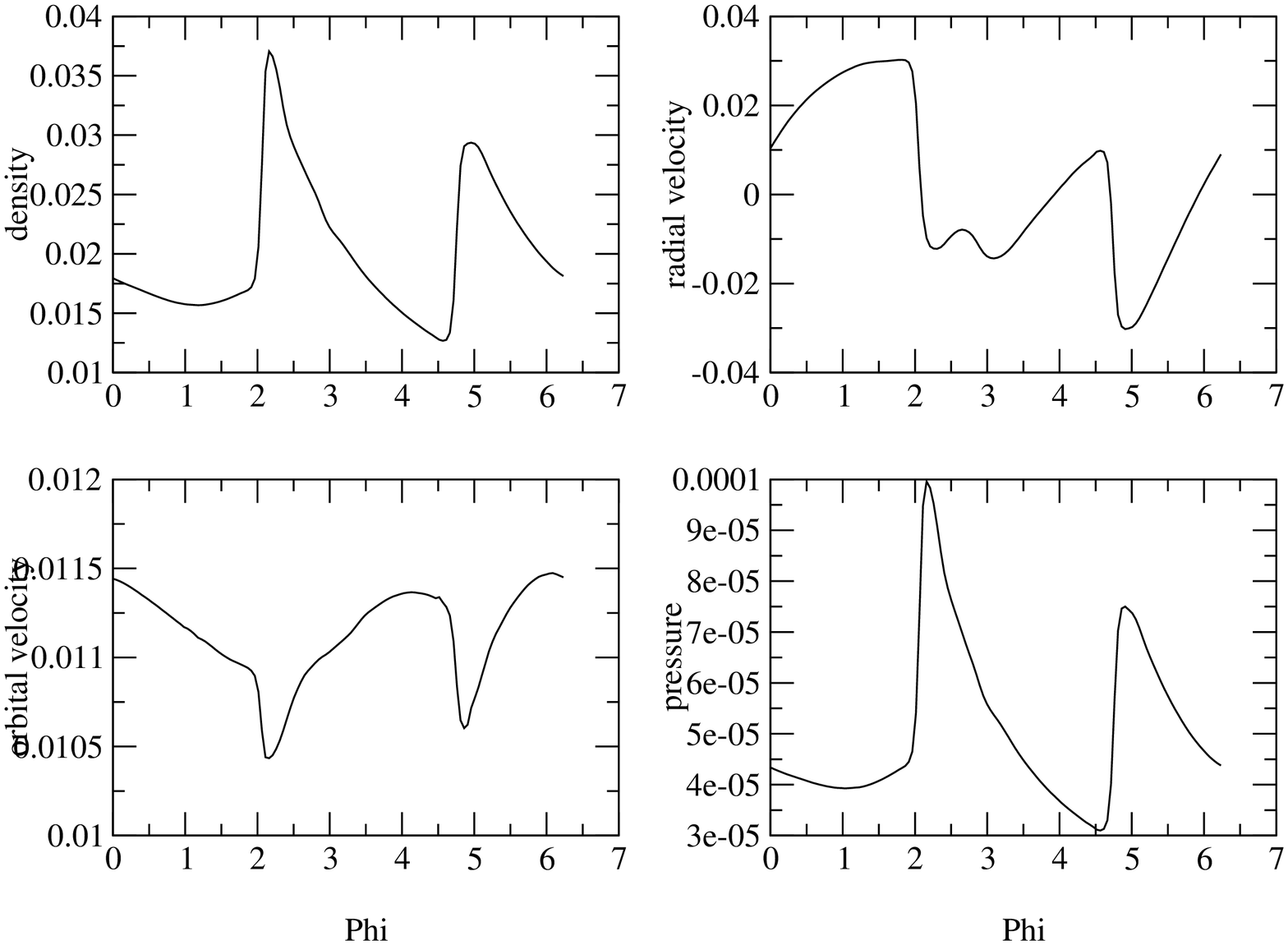}}
\caption{Plotting the density, radial velocity, orbital velocity and
pressure for accretion disk at $t=47526M$ in fixed $r=20.12M$.}  
\label{gamma=1.05_6}
\end{figure}
\end{center}

In order to watch the behavior of the spiral shock on an accretion 
disk while no matter gets into accretion disk, the injected gas is 
stooped. It is numerically observed that initial structure of 
two-armed spiral shock wave
is slightly changed because the hydrodynamical forces given by the
shock, which is created by injected gas, is gone. In order 
to balanced the forces on the accretion disk, the spiral shock waves 
are kept, which is more compact than the one in the beginning
of this simulations, around the black hole. This newly formed
two-armed spiral shocks are plotted at $t=47526M$ in
Fig.\ref{gamma=1.05_5}. These shocks are in the steady state and two-arms 
spiral shock
almost $180^o$ apart to each other. In Fig.\ref{gamma=1.05_6} we depict
$1D$ cut at fixed radial coordinate, $r=20.12M$ for density, radial
velocity, orbital velocity and pressure at $t=47526M$. Two-armed
spiral shocks are clearly observed. Angular momentum transform is also
seen in that graphic because the orbital velocities of spiral arms are
less than the orbital velocity of accretion disk.



\subsection{$\gamma=1.2$}
\label{gamma=1.2}

In this case, the same problem  is solved for accretion disk 
except  $\gamma=1.2$ which is more expressible than the $\gamma=1.05$. In
order to accrete an accretion disk, the same injection is made from outher 
boundary with the case for $\gamma=1.05$ and numerical simulation is run 
until solution reach to steady state. Afterall 
the injection is stooped and the behavior of accretion disk, and
spiral shock wave are numerically observed.

\begin{center}
\begin{figure}
\centerline{\epsfxsize=14.cm \epsfysize=14.cm
\epsffile{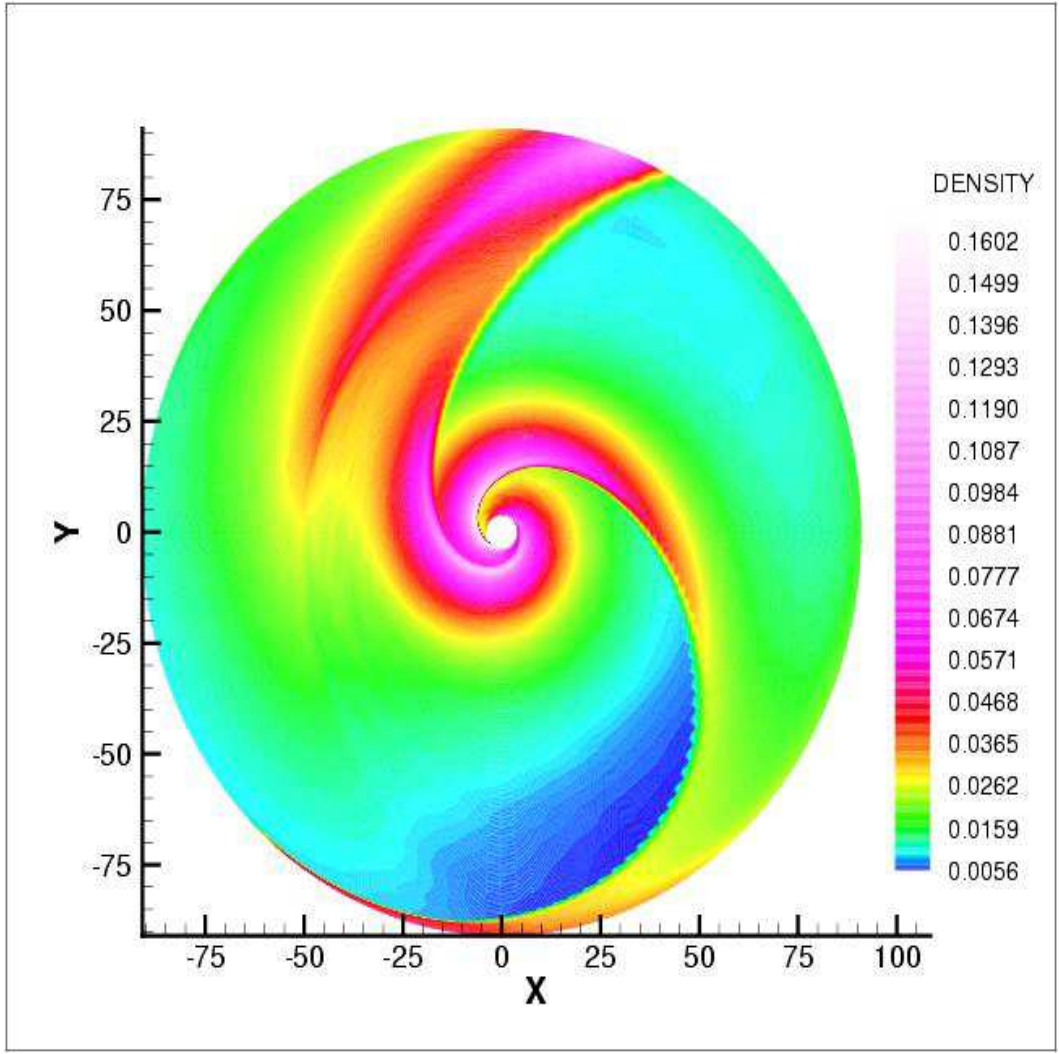}}
\caption{Plotting the density in the $r-\phi$ plane with color for
$\gamma=1.2$. It is taken at $t=17502M$. It is in
the steady state and  two-armed spiral shock wave is already created and
the structure of disk does not change.}
\label{gamma=1.2_2}
\end{figure}
\end{center}

We do injection from outer boundary continuously to create an accretion
disk around the black hole for $\gamma=1.2$. In Fig.\ref{gamma=1.2_2},
the two-armed spiral shock waves are created. The
solution is almost in steady state in Fig.\ref{gamma=1.2_2}. 
Eventhough the solution is not in steady
state, the structure of accretion disk is not change during the
evolutions. The result for $\gamma=1.2$ comparable with
Makita \emph{et al.} (2000). But there are some differences around the black hole
because our code fully relativistic and it can give detail structure
 when the matter close to black hole. Another differences in my code
and Makita \emph{et al.} (2000) is that we use inflow boundary condition that is
allow to gas fall into black hole but they use freezing boundary. The
using different boundary close to black hole also makes big difference
in the structure of spiral shock waves.

\begin{center}
\begin{figure}
\centerline{\epsfxsize=14.cm \epsfysize=14.cm
\epsffile{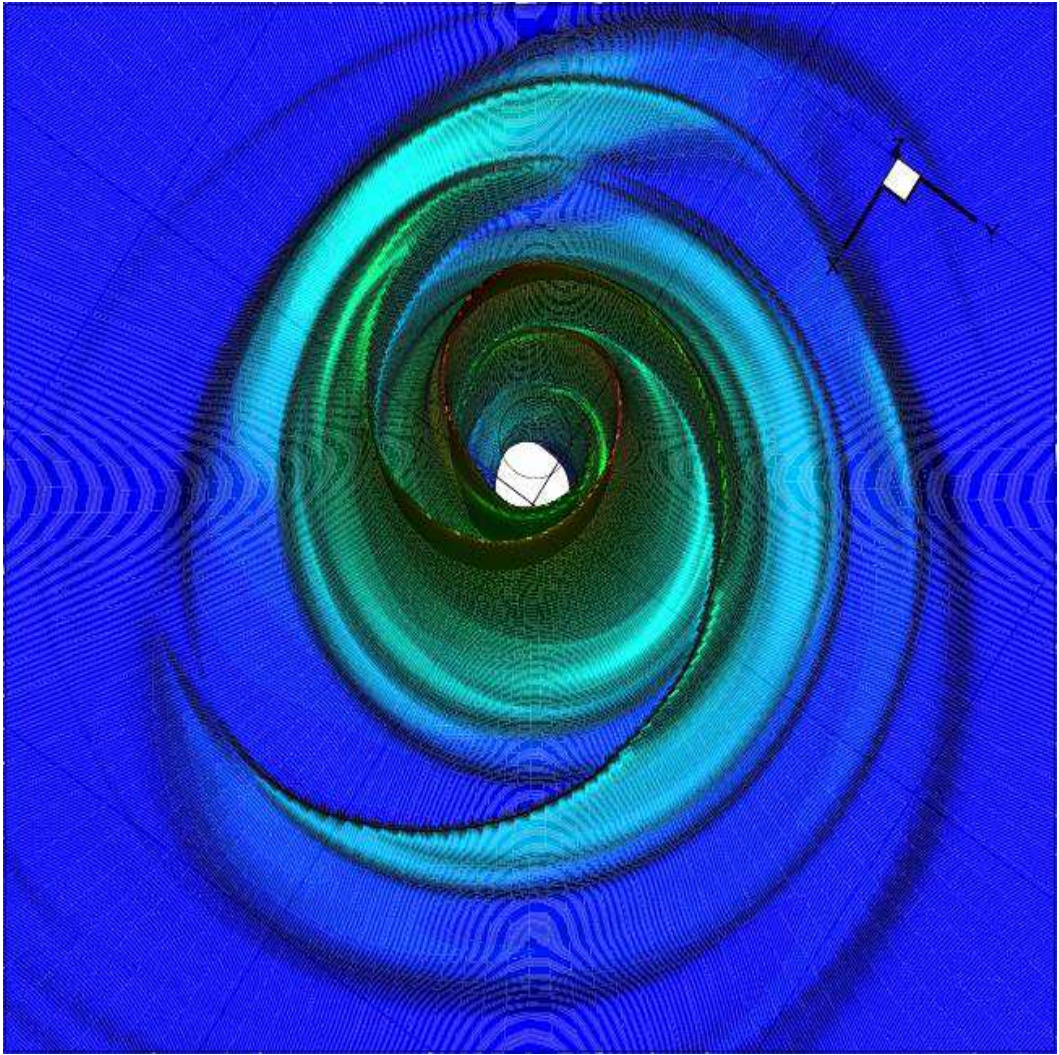}}
\caption{Plotting the density in the $r-\phi$ plane with color for
$\gamma=1.2$. It is taken at $t=19475M$ after injected is stopped
at$t=17502M$ . It is seen  that two-armed spiral shock wave is still
kept because they are created tidal forces on the accretion disk.}
\label{gamma=1.2_3}
\end{figure}
\end{center}

\begin{center}
\begin{figure}
\centerline{\epsfxsize=14.cm \epsfysize=14.cm
\epsffile{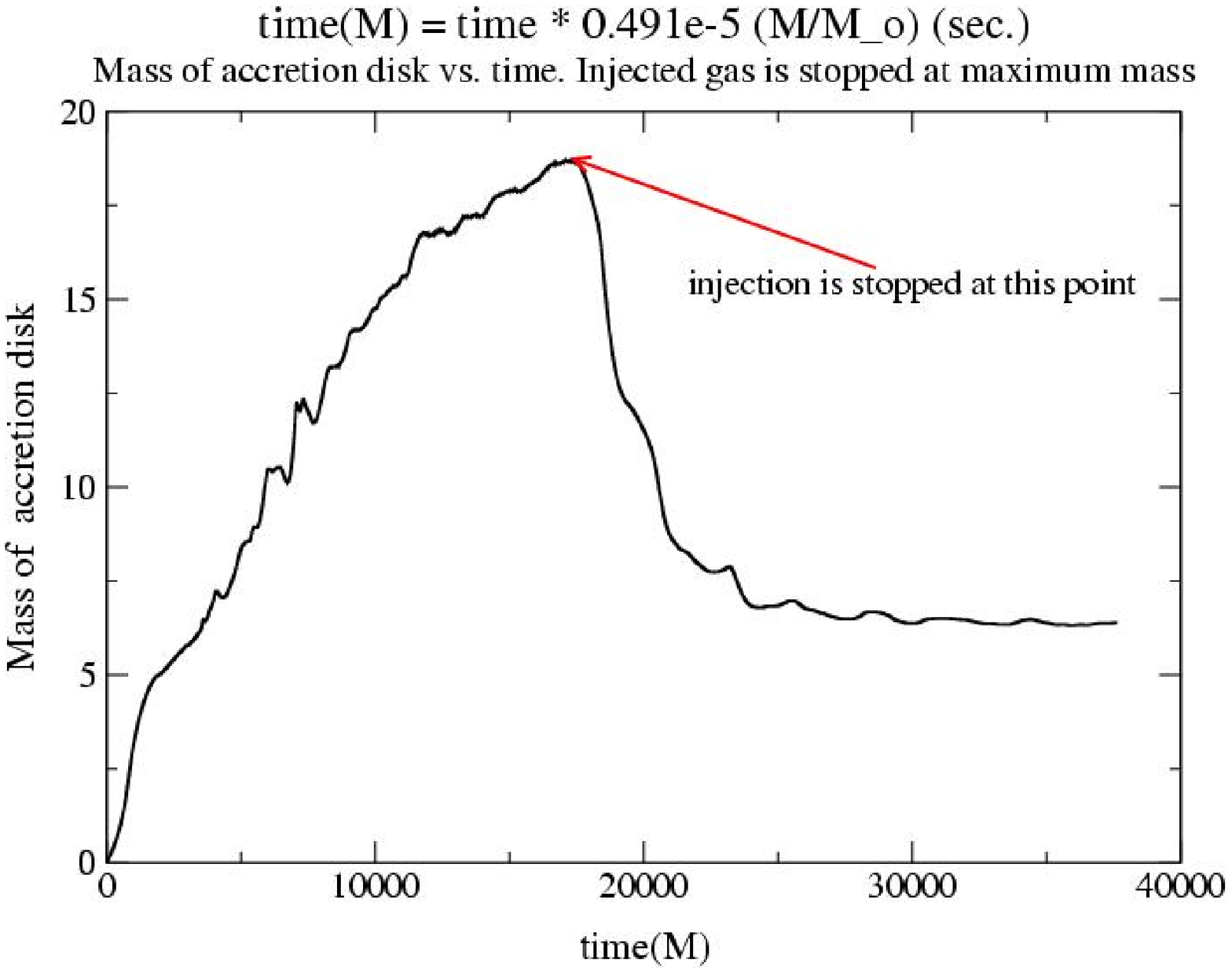}}
\caption{Mass of accretion disk vs. time is plotted during the hole
evolution. The injection is stopped at maximum mass.} 
\label{gamma=1.2_6}
\end{figure}
\end{center}

In order to understand the how spiral arms behave after companion star 
is removed, the injecting gas from outher boundary is stooped. 
After the injecting is not allowed any more at $t = 17502M$, 
Fig.\ref{gamma=1.2_3} displays numerical result 
at $t = 19475M$ that two-armed spiral shock waves are still 
kept because of tidal forces between companion star and black hole. Mass of 
the accretion  during the whole process is given at Fig.\ref{gamma=1.2_6}. 
It reaches a maximum mass, which is almost called steady-state point, 
whenever companion star is removed mass of accretion disk stars to decrease
and goes to steady-state point. That point has also two-armed spiral shock 
waves.

\section{Conclusion}
\label{Coclusion}

The spiral shock waves on an accretion disk are an important mechanism to
transform the angular momentum. In this simulation it is concluded
that accretion disk is rotating with sub-Keplerian velocity, which is
carrying positive angular momentum,  and spiral shock waves moving much
slower than the disk and they are carrying the negative angular
momentum. When the disk material hits to spiral wave, the angular
momentum of accretion disk is transform to out of accretion disk. The
shock waves around the black hole are a mechanism of transforming
gravitational energy to radiation energy which is observed by the
different $x-$ray  observatory satellite, such as Chandra.

Spiral arms in an accretion disk are one of the mechanism to emit $x-$ray.
We have looked at the spiral structure, which is created at around the black
hole when matter falls from companion star to primary star. Spiral structure
 in an accretion disk is created under certain condition. To create spiral 
arms adiabatic index must be less or equal than $1.2$.

In this simulations, three different simulations are done for $\gamma = 1.1, 1.2$
and $5/3$. These simulations show that we do not have any spiral arms for 
$\gamma =  5/3$ (the graphic for his case did not put here) but for the 
others. These results are consistent with results
from Makita \emph{et al.} (2000). We have also watched behavior of accretion disk after 
companion star stop injection. The two-armed spiral structure kept during  
the evolution.  It concludes that the spiral shocks are formed by tidal 
forces, not by the inflow, of which claim was posed by Bisikalo \emph{et al.}
(1998b).

From the point of view of dependence $\gamma$, it is concluded that spiral 
arms are 
more tightly in smaller $\gamma$ cases than larger ones. Lower  $\gamma$ means
cooler disk with larger Mach number of the flow. Our results are comparable 
with those of Makita \emph{et al.} (2000) who solve Newtonian 
hydrodynamical equation. Our and
their results are also agree while adiabatic index, $\gamma$, is bigger than 
$1.2$,  two armed spiral shock wave is not created which is observed 
for $\gamma = 5/3$.  Even the accretion disk may not be formed.

\begin{acknowledgments}
I would like to thank Joan M. Centrella, Cole Miller, Demos Kazanas 
and Tod Strohmayer for a useful discussion. This project was carried out 
at NASA/GSFC, Laboratory of High Energy Astrophysics.
It is supported by NASA/GSFC IR\&D. It has been performed using NASA super 
computers/T3E clusters.
\end{acknowledgments}

\vspace{2cm}

\appendix

\section{The Analytic Solution of Geodesics Flows}
\label{analytic solution for free fall}

Pressureless gas, also called  dust,  falling onto a black hole in the 
radial direction, called geodesic flow. It can be also called
free falling gas  because there are no pressure
forces opposing the inward motion of the gas. 

We use the fact that the elements
of the accreting fluid fall along geodesics to get the analytic
solution . In axisymmetric,
steady-state flows the binding energy per baryon $h U_t$ is
conserved. Hence for dust particles, $h=1$ and the gravitational binding energy
$U_t$ will remain constant. Since $U^{\mu} U_{\mu} = -1$, $v^r(r)$ is now
determined in terms of input an constant $U_t$ and the known metric
functions.  Note that $U_{\theta} = U_{\phi} = 0 $.

First, we start with the geodesic equation for free falling dust:

\begin{eqnarray}
\nabla_{\vec{u}} \vec{U} = 0. \nonumber \\
U^{\alpha}_{; \beta} U^{\beta} = (U^{\alpha}_{, \beta} +
\Gamma^{\alpha}_{\gamma \beta} U^{\gamma}) U^{\beta} = 0.
\label{geodesic1}
\end{eqnarray}

\noindent
Eq.(\ref{geodesic1}) can be rewritten  using  $U^{\alpha} =
dx^{\alpha}/d \lambda$,

\begin{eqnarray}
\frac{d U^{\alpha}}{d \lambda} + \Gamma^{\alpha}_{\gamma \beta}
U^{\gamma} U^{\beta} = 0.
\label{geodesic2}
\end{eqnarray}

\noindent
Substituting the index $\alpha = 0 = t$  in Eq.(\ref{geodesic2}) gives

\begin{eqnarray}
\frac{d U^{t}}{d \lambda} + \Gamma^{t}_{\gamma \beta}
U^{\gamma} U^{\beta} = 0.
\label{geodesic3}
\end{eqnarray}

\noindent
Most of the Christoffel symbols at Eq.(\ref{geodesic3}) are zero,
except $ \Gamma^{t}_{tr} =  \Gamma^{t}_{rt} = \frac{1}{2} 
g^{tt} \partial_r g_{tt} = \frac{1}{2} (2M/(r(r-2M)))$. The
substituting the non-zero Christoffel symbols into 
Eq.(\ref{geodesic3}) gives us

\begin{eqnarray}
\frac{d U^{t}}{d r} = - \frac{2M}{r(r-2M)} U^t,
\label{geodesic4}
\end{eqnarray}

\noindent
where $dU^t/d \lambda = (dU^t/d r) (dr/d \lambda)$.

\noindent
After doing some straightforward integration, Eq.(\ref{geodesic4})
goes to 

\begin{eqnarray}
U^t = \frac{1}{(1 - \frac{2M}{r})} .
\label{geodesic5}
\end{eqnarray}

To compute the radial component of the fluid velocity of geodesic gas,
we need to know the radial 
four velocity of the gas. In order to compute that, we  use the
normalization of four velocities, which is

\begin{eqnarray}
U^{\mu} U_{\mu} = -1 \nonumber \\
U^t \gamma_{tt} U^t + U^r \gamma_{rr} \gamma^r = -1.
\label{geodesic6}
\end{eqnarray}

\noindent
We substitute  Eq.(\ref{geodesic5}) into Eq.(\ref{geodesic6}), to
\begin{eqnarray}
U^r = \sqrt{\frac{2M}{r}}.
\label{geodesic7}
\end{eqnarray}

\noindent
Using the relations between the four and three velocities, $U^r = W
v^r$, $U^t = W/\sqrt{1- (2M/r)}$, we get

\begin{eqnarray}
v^r = \sqrt{\frac{2M}{r}} \sqrt{1 - \frac{2M}{r}},
\label{geodesic8}
\end{eqnarray}

\noindent
where $v^r$ is the velocity which is observed by an observer outside the horizon.

Now, we compute the density  using the
continuity equation from Eq.(\ref{desired equation}) which is 

\begin{eqnarray}
\partial_t (\sqrt{\gamma} W \rho) + \partial_i (\alpha v^r D) = 0.
\label{Equation of density1}
\end{eqnarray}

\noindent
Since we are looking for a steady state solution the time derivative
of variables is zero and we have 
only the radial derivative in Eq.(\ref{Equation of density1}). Then the
density equation becomes
 
\begin{eqnarray} 
\partial_i (\alpha v^r D) = 0.
\label{Equation of density2}
\end{eqnarray}

\noindent
After doing integration of Eq.(\ref{Equation of density2}), we get

\begin{eqnarray} 
\alpha v^r D = d,
\label{Equation of density3}
\end{eqnarray}

\noindent
where $d$ is an integration constant. Now $D$ can be computed  from
Eqs.(\ref{geodesic8}) and (\ref{Equation of density3}), 

\begin{eqnarray} 
D = \frac{d}{(1 - \frac{2 M}{r}) (\frac{2 M}{r})^{\frac{1}{2}}}.
\label{conserved density for free fall}
\end{eqnarray}

\noindent
Finally, we compute the density $\rho$ from Eqs.(\ref{geodesic8}) and
(\ref{conserved density for free fall})

\begin{eqnarray}
\rho = \frac{1}{W} \frac{d}{r^2 (\frac{2 M}{r})^{\frac{1}{2}} (1 - \frac{2
M}{r})^{\frac{1}{2}}},
\label{density for free fall}
\end{eqnarray}

\noindent
where $W$ is the Lorentz factor and given by

\begin{eqnarray}
W = \frac{1}{(1 - \frac{2 M}{r})^{\frac{1}{2}}}.
\label{lorentz factor for free fall}
\end{eqnarray}


\section{The Analytic Representation of Circular Motion of a Test
Particle} 
\label{The Analytic Representation of Circular Motion of The Test Particle}

In this Appendix we compute the angular velocity and circular velocity
of a particle on a circular orbit in the Schwarzschild spacetime, 
analytically. 

Since Schwarzschild geometry is time independent and spherically
symmetric, the conserved quantities can be determined by the trajectory of
particles. Because of spherical symmetry, motion is always defined in at
a single plane and we can choose this plane to be the equatorial plane
$(\theta = \pi/2)$. Then $\theta$ is constant in that plane for the motion
of particles  and the $\theta$ derivatives  
vanish. The components of the momentum (Schutz \emph{et al.} 1985) are

\begin{eqnarray}
p^t = g^{tt}p_t = \frac{m E}{(1- \frac{2M}{r})} \nonumber \\
p^r = m \frac{dr}{d \tau} \nonumber \\
p^{\theta} = 0 \\
p^{\phi} = g^{\phi \phi}p_{\phi} = \frac{m L}{r^2}, \nonumber
\label{momentums for circular motion}
\end{eqnarray}

\noindent
where $m$, $E$ , $\tau$ and $L$ are the mass of particle, total energy,
proper time and angular momentum, respectively. Here $E = -p_t/m$ and
$L = p_{\phi}/m$. 

Now, we can derive the equation of motion for a  particle in the  equatorial
plane using Eq.(\ref{momentums for circular motion}) and the conservation
relation, $\vec{p} . \vec{p} = -m^2$. This gives us

\begin{eqnarray}
\left(\frac{dr}{d \tau}\right)^2 = E^2 - \left(1 - \frac{2M}{r}\right) 
\left(1 + \frac{L^2}{r^2}\right).
\label{equation for circular motion}
\end{eqnarray}

\noindent
Eq.(\ref{equation for circular motion}) can be rewritten by defining
an effective potential $V(r)$ and we get

\begin{eqnarray}
\left(\frac{dr}{d \tau}\right)^2 = E^2 - V^2(r),
\label{equation for circular motion2}
\end{eqnarray}

\noindent
where $V^2(r) = (1 - \frac{2M}{r}) (1 + \frac{L^2}{r^2})$.

\noindent
Eq.(\ref{equation for circular motion2}) implies that since the left side of
that equation is positive or zero, the total energy of a  trajectory can
be bigger or equal to the effective potential.

In order to compute the angular velocity and period of a particle in a
circular orbit, we differentiate  Eq.(\ref{equation for circular
motion2}) with respect to $\tau$, and get

\begin{eqnarray}
\frac{d^2 r}{d \tau^2} = -\frac{1}{2} \frac{d V^2(r)}{dr}.
\label{equation for circular motion3}
\end{eqnarray}

\noindent
It is clear from Eq.(\ref{equation for circular motion3}) that a
circular orbit, which has constant $r$, is possible only at a minimum
or maximum of the effective potential, $V^2(r)$. In a circular orbit
$r$ is constant  and the left side of Eq.(\ref{equation for circular
motion3}) goes to zero. If we take  $d^{2}r/d \tau^2 = 0$ and
substitute in the expression for the effective potential, we can compute 
the circular orbit radius as 

\begin{eqnarray}
r = \frac{L^2}{2M} \left(1 \pm \sqrt{\left(1 - \frac{12M^2}{L^2}\right)}\right).
\label{equation for circular motion4}
\end{eqnarray}

\noindent
From  Eq.(\ref{equation for circular motion4}), a stable circular
orbit at radius $r$ has angular momentum which is 

\begin{eqnarray}
L^2  = \frac{Mr}{1 - \frac{3M}{r}}.
\label{equation for circular motion5}
\end{eqnarray}

\noindent
The total energy in a circular orbit is $E^2 = V^2$ and it is

\begin{eqnarray}
E = \left(1 - \frac{2M}{r}\right)^2/\left(1 - \frac{3M}{r}\right).
\label{equation for circular motion6}
\end{eqnarray}

Now, the non zero components of the four velocity of a particle in the plane 
are

\begin{eqnarray}
\frac{d\phi}{d\tau} = U^{\phi} = \frac{P^{\phi}}{m} = g^{\phi \phi}
\frac{P_{\phi}}{m} = g^{\phi \phi} L = \frac{L}{r^2} 
\label{equation for circular motion7}
\end{eqnarray}

\noindent
and

\begin{eqnarray}
\frac{dt}{d\tau} = U^t = \frac{P^t}{m} = g^{tt}\frac{P_t}{m} = g^{t
t} (-E) = \frac{E}{1 - \frac{2 M}{r}}.
\label{equation for circular motion8}
\end{eqnarray}

\noindent
We find the angular velocity by dividing Eq.(\ref{equation for
circular motion7}) by Eq.(\ref{equation for circular motion8}):

\begin{eqnarray}
\frac{d \phi}{d t}  = \frac{d \phi/d \tau}{dt/d \tau} = \sqrt{\frac{M}{r^3}};
\label{equation for circular motion9}
\end{eqnarray}

\noindent
this is called the Keplerian angular velocity.

\noindent
Finally, we can compute the circular velocity of a particle using the
definition of four velocity relativistic hydrodynamical equations given 
in Section \ref{formulation}
, Eq.(\ref{equation for circular motion7}) and Eq.(\ref{equation for
circular motion8}). We get

\begin{eqnarray}
v^{\phi} = \frac{1}{\alpha} \frac{U^{\phi}}{U^t} \nonumber \\
 \nonumber \\
v^{\phi} =\frac{1}{\sqrt{(1 - \frac{2M}{r})}} \sqrt{\frac{M}{r^3}}.
\label{equation for circular motion10}
\end{eqnarray}


\vspace{2cm}

\begin{center}
{\bf References}
\end{center}

\noindent
Bisikalo, D. V., Boyarchuk, A. A., Chechetin, V. M., 
Kuznetsov, O. A., Molteni D.. 1998b, MNRAS. \\
Chakabarti, S. K., Molteni, D. 1993, Astrophys. J. {\bf
417}:671-676. \\
Chakabarti, S. K. 1997, Astrophys. J. {\bf 484}:313-322. \\
Dgani, R., Livio, M., Regev, O. 1994, Astrophys. J., 436, 270.  \\
Donat, R., Font, J. A., Ibanez, J. M., and Marquina, A. 1998,
J.Comput. Phys. {\bf 146}, 58. \\
D\"{o}nmez, O, submitted to Astophysics and Space Science(2003) \\
Font, J. A., Miller, M., Suen, W.-M., Tobias, M. 2000,
Physical Review D, {\bf 61}, 044011. \\
Godon, P. 1997, Astrophys. J., 480, 329. \\
Hawley, J. F., Smarr, L., Wilson, J. R. 1985, Astrophys. J. {\bf
277}: 296-311.   \\
Lanzafame, G., Molteni, Chakabarti, D.,S. K. 1998,
Mon. Not. R. Astron. Soc. {\bf 299}, 799-804. \\
Makita, M., Miyawaki, K. and Matsuda, T. 2000,
Mon. Not. R. Astron. Soc. {\bf 000}, 1,13. \\
Molteni, D., Toth, G., Kuznetsov, O. A. 1999, Astrophys. J. {\bf
516}:411-419. \\
Rozyczka, M., Spruit, H. C. 1993, Astrophys. J., 417, 677. \\
Schutz, B. F. 1985, in \emph{A First Course in General
Relativity}, edited by B. F. Schutz  (Cambridge University Press,
Cambridge). \\ 
Steeghs, D., Harlaftis, E. T., Horne, K. 1997, MNRAS, {\bf 209},
L28.


\newpage

\newpage

\newpage

\newpage

\newpage

\newpage

\newpage

\newpage

\newpage

\newpage

\newpage

\end{document}